\documentclass[envcountsect]{llncs}

\spnewtheorem{wfcond}{Well-formedness conditions}[section]{\bf}{\textnormal}

\begin{document}

\title{Quantum Pushdown Automata}

\author{Marats Golovkins}

\institute{
 Institute of Mathematics and Computer Science,
 University of Latvia, Rai\c na bulv. 29, Riga, Latvia\thanks{%
Research partially supported by the Latvian Council of Science,
grant 96-0282 and grant for PhD students; European Comission, contract IST-1999-11234;
Swedish Institute, project ML2000
}
\\ \email{marats@latnet.lv}
}

\maketitle

\begin{abstract}
Quantum finite automata, as well as quantum pushdown automata (QPA) were first
introduced by C. Moore, J. P. Crutchfield \cite{MC 97}.
In this paper we introduce the notion of QPA in a
non-equivalent way, including unitarity criteria, by
using the definition of quantum finite automata of \cite{KW 97}.
It is established that the unitarity criteria of QPA are not equivalent
to the corresponding unitarity criteria of quantum Turing machines
\cite{BV 97}. We show that QPA can recognize every regular language.
Finally we present some simple languages recognized by QPA, not
recognizable by deterministic pushdown automata.
\end{abstract}

\section{Introduction}
Nobel prize winner physicist R. Feynman asked in 1982, what effects may
have the principles of quantum mechanics on computation \cite{Fe 82}. He
gave arguments that it may require exponential time to simulate quantum
mechanical processes on classical computers. This served as a basis to
the opinion that quantum computers may have advantages versus classical
ones. It was in 1985, when D. Deutsch introduced the notion of quantum
Turing machine \cite{De 85} and proved that quantum Turing machines compute
the same recursive functions as classical deterministic Turing machines
do. P. Shor discovered that by use of quantum algorithms it is possible to 
factorize large integers and compute discrete logarithms in a polynomial
time \cite{Sh 94}, what resulted into additional interest in quantum 
computing and attempts to create quantum computers. First steps have
been made to this direction, and first quantum computers which memory is
limited by a few quantum bits have been constructed \cite{KLMT 99}. To make
quantum computers with larger memory feasible, one of the problems is to
minimize error possibilities in quantum bits. Quantum error correction 
methods are developed \cite{CRSS 98} which would enable quantum
computers with larger quantum memory.

For the analysis of the current situation in quantum computation and information
processing and main open issues one could see \cite{Gr 99}.

Quantum mechanics differs from the classical physics substantially. It
is enough to mention {\em Heisenberg's uncertainty principle}, which states
that it is impossible to get information about different parameters of
quantum particle simultaneously precisely. Another well known distinction
is the impossibility to observe quantum object without changing it.

Fundamental concept of quantum information theory is {\em quantum bit}.
Classical information theory is based on classical bit, which has two
states $0$ and $1$. The next step is {\em probabilistic bit}, which
can be $0$ with probability $\alpha$ and $1$ with probability $\beta$,
where $\alpha+\beta=1$. Quantum bit or {\em qbit} is similar to
probabilistic bit with the difference that $\alpha$ and $\beta$ are complex
numbers with the property $|\alpha|^2+|\beta|^2=1$. It is common to denote
qbit as $\alpha|0\rangle+\beta|1\rangle$. As a result of {\em measurement},
we get $0$ with probability $|\alpha|^2$ and $1$ with probability
$|\beta|^2$.

Every computation done on qbits is accomplished by means of unitary
operators. Informally, every unitary operator can be interpreted as a
evolution in complex space. Therefore one of the basic properties of
unitary operators is that every quantum computing process not disturbed
by measurements is reversible. Unitarity is rather hard requirement which
complicates programming of quantum devices.
The following features of quantum computers are most important:
\begin{enumerate}
\item Information is represented by qbits.
\item Any step of computation can be represented as a unitary operation,
therefore computation is reversible.
\item Quantum information cannot be copied.
\item Quantum parallelism; quantum computer can compute several paths
simultaneously, however as a result of measurement it is possible to get the
results of only one computation path.
\end{enumerate}

Opposite to quantum Turing machines, quantum finite automata (QFA) represent
the finite model of quantum computation. QFA were first introduced by 
\cite{MC 97} (measure-once QFA), which were followed by a more elaborated
model of \cite{KW 97} (measure-many quantum finite automata). Since then
QFA have been studied a lot, various properties of these automata are
considered in \cite{ABFK 99,AF 98,BP 99,V 00}. Quantum finite one counter
automata were introduced by \cite{Kr 99}.

The purpose of this paper is to introduce
a quantum counterpart of pushdown automata, the next most important model
after finite automata and Turing machines.
The first definition of quantum pushdown automata was suggested by
\cite{MC 97}, but here the authors actually deal with the so-called
generalized quantum pushdown automata, which evolution does not have to be
unitary. However a basic postulate of quantum
mechanics imposes a strong constraint on any quantum machine model: it has to
be unitary, otherwise it is questionable whether we can speak
about {\em quantum} machine.
That's why it was considered necessary to re-introduce quantum pushdown
automata by
giving a definition which would conform unitarity requirement. Such
definition would enable us to study the properties of quantum pushdown
automata.

The following notations will be used further in the paper:\\
$z^*$ is the complex conjugate of a complex number $z$.\\
$U^*$ is the Hermitian conjugate of a matrix $U$.\\
$I$ is the identity matrix.\\
$\varepsilon$ is empty word.

\begin{definition}
\label{def1}
Matrix $U$ is called {\em unitary}, if $UU^*=U^*U=I$.
\end{definition}

If $U$ is a finite matrix, then $UU^*=I$ iff $U^*U=I$. However this is not
true for infinite matrices:

\begin{example}
$$
U=\left(
\begin{array}{cccccc}
\frac{1}{\sqrt 2} & 0 & 0 & 0 & 0 & \dots\\
\frac{1}{\sqrt 2} & 0 & 0 & 0 & 0 & \dots\\
0 & 1 & 0 & 0 & 0 & \dots\\
0 & 0 & 1 & 0 & 0 & \dots\\
0 & 0 & 0 & 1 & 0 & \dots\\
\vdots & \vdots & \vdots & \vdots & \vdots & \ddots
\end{array}
\right)
$$
Here $U^*U=I$ but $UU^*\neq I$.
\end{example}

\begin{lemma}
\label{asst}
If infinite matrices $A,B,C$ have finite number of nonzero elements
in each row and column, then their multiplication is associative:
$(AB)C=A(BC)$.
\end{lemma}
\begin{proof}
The element of matrix $(AB)C$ in $i$-th row and $j$-th column is
$k_{ij}=\sum\limits_{s=1}^\infty\sum\limits_{r=1}^\infty a_{ir}b_{rs}c_{sj}$.
The element of matrix $A(BC)$ in the same row and column is
$l_{ij}=\sum\limits_{r=1}^\infty\sum\limits_{s=1}^\infty a_{ir}b_{rs}c_{sj}$.
As in the each row and column of matrices $A,B,C$ there is a finite number of
nonzero elements, it is also finite in the given series. Therefore the
elements of the series can be rearranged, and $k_{ij}=l_{ij}$.\qed
\end{proof}

As noted further in the paper infinite matrices with finite number of
nonzero elements in each row and column describe the work of pushdown
automata. Further lemmas state some properties of such
matrices.

\begin{lemma}
\label{th1}
If $U^*U=I$, then the norm of any row in the matrix $U$ does not exceed $1$.
\end{lemma}
\begin{proof}
Let us consider the matrix $S=UU^*$. The element of this matrix
$s_{ij}=\langle r_j|r_i\rangle$, where $r_i$ is $i$-th row of the matrix
$U$. Let us consider the matrix $T=S^2$. The diagonal element of this
matrix is
$$
t_{ii}=\sum\limits_{k=1}^\infty s_{ik}s_{ki}=\sum\limits_{k=1}^\infty
\langle r_k|r_i\rangle\langle r_i|r_k\rangle=\sum\limits_{k=1}^\infty
|\langle r_k|r_i\rangle|^2.
$$
On the other hand, taking into account Lemma \ref{asst}, we get that
$$
T=S^2=(UU^*)(UU^*)=U(U^*U)U^*=UU^*=S.
$$
Therefore $t_{ii}=s_{ii}=\langle r_i|r_i\rangle$. It means that
\begin{equation}
\label{eq1}
\sum\limits_{k=1}^\infty|\langle r_k|r_i\rangle|^2=\langle r_i|r_i\rangle.
\end{equation}
This implies that every element of series (\ref{eq1}) does not exceed
$\langle r_i|r_i\rangle$. Hence 
$|\langle r_i|r_i\rangle|^2=\langle r_i|r_i\rangle^2\leq\langle r_i|r_i\rangle$.
The last inequality implies that $0\leq\langle r_i|r_i\rangle\leq 1$.
Therefore $|r_i|\leq 1$.\qed
\end{proof}

\begin{lemma}
\label{th2}
Let us assume that $U^*U=I$. Then the rows of the matrix $U$ are
orthogonal iff every row of the matrix has norm $0$ or $1$.
\end{lemma}
\begin{proof}
Let us assume that the rows of the matrix $U$ are orthogonal. Let
us consider equation (\ref{eq1}) from the proof of Lemma \ref{th1},
i.e.,
$\sum\limits_{k=1}^\infty|\langle r_k|r_i\rangle|^2=\langle r_i|r_i\rangle$.
As the rows of the matrix $U$ are orthogonal,
$\sum\limits_{k=1}^\infty|\langle r_k|r_i\rangle|^2=|\langle r_i|r_i\rangle|^2$.
Hence $\langle r_i|r_i\rangle^2=\langle r_i|r_i\rangle$, i.e.,
$\langle r_i|r_i\rangle=0$ or $\langle r_i|r_i\rangle=1$. Therefore
$|r_i|=0$ or $|r_i|=1$.

Let as assume that every row of the matrix has norm $0$ or $1$. Then
$\langle r_i|r_i\rangle^2=\langle r_i|r_i\rangle$ and in compliance
with the equation (\ref{eq1}),
$\sum\limits_{k\in\bbbn^+\setminus\{i\}}|\langle r_k|r_i\rangle|^2=0$.
This implies that $\forall k\neq i\ |\langle r_k|r_i\rangle|=0$.
Hence the rows of the matrix are orthogonal.\qed
\end{proof}

\begin{lemma}
\label{th3}
The matrix $U$ is unitary iff $U^*U=I$ and its rows are normalized.
\end{lemma}
\begin{proof}
Let us assume that the matrix $U$ is unitary. Then in compliance with
Definition \ref{def1}, $U^*U=I$ and $UU^*=I$, i.e, the rows of the
matrix are orthonormal.

Let us assume that $U^*U=I$ and the rows of the matrix are
normalized. Then in compliance with Lemma \ref{th2} the rows
of the matrix are orthogonal. Hence $UU^*=I$ and the matrix is
unitary.\qed
\end{proof}

This result is very similar to Lemma 1 of \cite{DS 96}.

\section{Quantum pushdown automata}

\begin{definition}
\label{def2}
A quantum pushdown automaton (QPA)\\
$A=(Q,\Sigma,T,q_0,Q_a,Q_r,\delta)$ is specified by a finite set of
states $Q$, a finite input alphabet $\Sigma$ and a stack alphabet
$T$, an initial state $q_0\in Q$, sets $Q_a\subset Q$,
$Q_r\subset Q$ of accepting and rejecting states, respectively, with 
$Q_a\cap Q_r=\emptyset$, and a transition function
$$
\delta:Q\times\Gamma\times\Delta\times Q\times\{\downarrow,\to\}\times
\Delta^*\longrightarrow\bbbc_{[0,1]},
$$
where $\Gamma=\Sigma\cup\{\#,\$\}$ is the input tape alphabet of $A$ and
$\#,\$$ are end-markers not in $\Sigma$, $\Delta=T\cup\{Z_0\}$ is the
working stack alphabet of $A$ and $Z_0\notin T$ is the stack base symbol;
$\{\downarrow,\to\}$ is the set of directions of input tape head.
The automaton must satisfy {\em conditions of well-formedness}, which
will be expressed below. Furthermore, the transition function is restricted
to a following requirement:

\begin{enumerate}
If $\delta(q,\alpha,\beta,q',d,\omega)\neq0$, then
\item $|\omega|\leq 2$;
\item if $|\omega|=2$, then $\omega_1=\beta$;
\item if $\beta=Z_0$, then $\omega\in Z_0 T^*$;
\item if $\beta\neq Z_0$, then $\omega\in T^*$.
\end{enumerate}
\end{definition}

Definition \ref{def2} utilizes that of classical pushdown automata
from \cite{Gu 89}.

\begin{wfcond}
\label{wfc1}
\begin{enumerate}
\item Local probability condition.
\begin{eqnarray}
&&\forall(q_1,\sigma_1,\tau_1)\in Q\times\Gamma\times\Delta\nonumber\\
\label{lpc1}
&&\sum\limits_{(q,d,\omega)\in Q\times\{\downarrow,\to\}\times\Delta^*} 
|\delta(q_1,\sigma_1,\tau_1,q,d,\omega)|^2=1.
\end{eqnarray}
\item Orthogonality of column vectors condition.
\begin{eqnarray}
&&\mbox{For all triples }(q_1,\sigma_1,\tau_1)\neq(q_2,\sigma_1,\tau_2)
\mbox{ in }Q\times\Gamma\times\Delta\nonumber\\
\label{ocv1}
&&\sum\limits_{(q,d,\omega)\in Q\times\{\downarrow,\to\}\times\Delta^*}
\delta^*(q_1,\sigma_1,\tau_1,q,d,\omega)\delta(q_2,\sigma_1,\tau_2,q,d,\omega)=0.
\end{eqnarray}
\item Row vectors norm condition.
\begin{eqnarray}
&&\forall(q_1,\sigma_1,\sigma_2,\tau_1,\tau_2)\in Q\times\Gamma^2\times\Delta^2
\ \ \ \ \ \ \ \ \ \ \ \ \ \ \ \ \ \ \ \ \ \ \ \ \ \ \ \ \ \ \ \ \ \ \ \ \ \ \ \ \ \ \ \ \ \ \ \ \ \ \ \ \ \nonumber\\
\label{rvn1}
&&\sum\limits_{(q,\tau,\omega)\in Q\times\Delta\times\{\varepsilon,\tau_2,\tau_1\tau_2\}}
|\delta(q,\sigma_1,\tau,q_1,\to,\omega)|^2+|\delta(q,\sigma_2,\tau,q_1,\downarrow,\omega)|^2=1.
\end{eqnarray}
\item Separability condition I.
\begin{eqnarray}
&&\forall(q_1,\sigma_1,\tau_1),(q_2,\sigma_1,\tau_2)\in Q\times\Gamma\times\Delta,
\ \forall\tau_3\in\Delta\nonumber\\
\label{s1a1}
&&a)\ \sum\limits_{(q,d,\tau)\in Q\times\{\downarrow,\to\}\times\Delta}
\delta^*(q_1,\sigma_1,\tau_1,q,d,\tau)\delta(q_2,\sigma_1,\tau_2,q,d,\tau_3\tau)+{}\nonumber\\
&&{}+\sum\limits_{(q,d)\in Q\times\{\downarrow,\to\}}
\delta^*(q_1,\sigma_1,\tau_1,q,d,\varepsilon)\delta(q_2,\sigma_1,\tau_2,q,d,\tau_3)=0;\\
\label{s1b1}
&&b)\ \sum\limits_{(q,d)\in Q\times\{\downarrow,\to\}}
\delta^*(q_1,\sigma_1,\tau_1,q,d,\varepsilon)\delta(q_2,\sigma_1,\tau_2,q,d,\tau_2\tau_3)=0.
\end{eqnarray}
\item Separability condition II.
\begin{eqnarray}
&&\forall(q_1,\sigma_1,\tau_1),(q_2,\sigma_2,\tau_2)\in Q\times\Gamma\times\Delta\nonumber\\
\label{s21}
&&\sum\limits_{(q,\omega)\in Q\times\Delta^*}
\delta^*(q_1,\sigma_1,\tau_1,q,\downarrow,\omega)\delta(q_2,\sigma_2,\tau_2,q,\to,\omega)=0.
\end{eqnarray}
\item Separability condition III.
\begin{eqnarray}
&&\forall(q_1,\sigma_1,\tau_1),(q_2,\sigma_2,\tau_2)\in Q\times\Gamma\times\Delta,
\ \forall\tau_3\in\Delta,\ \forall d_1,d_2\in\{\downarrow,\to\},\ d_1\neq d_2\nonumber\\
\label{s3a1}
&&a)\ \sum\limits_{(q,\tau)\in Q\times\Delta}
\delta^*(q_1,\sigma_1,\tau_1,q,d_1,\tau)\delta(q_2,\sigma_2,\tau_2,q,d_2,\tau_3\tau)+{}\nonumber\\
&&{}+\sum\limits_{q\in Q}
\delta^*(q_1,\sigma_1,\tau_1,q,d_1,\varepsilon)\delta(q_2,\sigma_2,\tau_2,q,d_2,\tau_3)=0;\\
\label{s3b1}
&&b)\ \sum\limits_{q\in Q}
\delta^*(q_1,\sigma_1,\tau_1,q,d_1,\varepsilon)\delta(q_2,\sigma_2,\tau_2,q,d_2,\tau_2\tau_3)=0.
\end{eqnarray}
\end{enumerate}
\end{wfcond}

Let us assume that an automaton is in a state $q$, its input tape head
is above a symbol $\alpha$ and the stack head is above a symbol $\beta$.
Then the automaton undertakes following actions with an amplitude
$\delta(q,\alpha,\beta,q',d,\omega)$:
\begin{enumerate}
\item goes into the state $q'$;
\item if $d=`\to\textnormal{'}$, moves the input tape head one cell forward;
\item takes out of the stack the symbol $\beta$ (deletes it and moves
the stack head one cell backwards);
\item starting with the first empty cell, puts into the stack the string
$\omega$, moving the stack head $|\omega|$ cells forward.
\end{enumerate}

\begin{definition}
The {\em configuration} of a pushdown automaton is a pair
$|c\rangle=|\nu_i q_j\nu_k,\omega_l\rangle$, where the automaton is in a
state $q_j\in Q$, $\nu_i\nu_k\in\#\Sigma^*\$$ is a finite word on the input
tape, $\omega_l\in Z_0T^*$ is a finite word on the stack tape, the input
tape head is above the first symbol of the word $\nu_k$ and the stack head
is above the last symbol of the word $\omega_l$.
\end{definition}

We shall denote by $C$ the set of all configurations of a pushdown
automaton. The set $C$ is countably infinite. Every configuration 
$|c\rangle$ denotes a basis vector in the space $H_A=l_2(C)$.
Therefore a global state of $A$ in the space $H_A$ has a form
$|\psi\rangle=\sum\limits_{c\in C}\alpha_c|c\rangle$, where
$\sum\limits_{c\in C}|\alpha_c|^2=1$ and $\alpha_c\in\bbbc$ denotes the
amplitude of a configuration $|c\rangle$. If an automaton is in its global
state (superposition) $|\psi\rangle$, then its further step is equivalent to
the application of a linear operator (evolution) $U_A$ over the space $H_A$.

\begin{definition}
A linear operator $U_A$ is defined as follows:
$$
U_A|\psi\rangle=\sum\limits_{c\in C}\alpha_c U_A|c\rangle.
$$
If a configuration $c=|\nu_i q_j\sigma\nu_k,\omega_l\tau\rangle$, then
$$
U_A|c\rangle=\sum\limits_{(q,d,\omega)\in Q\times\{\downarrow,\to\}\times\Delta^*}
\delta(q_j,\sigma,\tau,q,d,\omega)|f(|c\rangle,d,q),\omega_l\omega\rangle,
$$
where
$$
f(|\nu_i q_j\sigma\nu_k,\omega_l\tau\rangle,d,q)=
\left\{
\begin{array}{l}
\nu_i q\sigma\nu_k,\mbox{if }d=`\downarrow\textnormal{'}\\
\nu_i\sigma q\nu_k,\mbox{if }d=`\to\textnormal{'}.
\end{array}
\right.
$$
\end{definition}

\begin{remark}
Although a QPA evolution operator matrix is infinite, it has a finite
number of nonzero elements in each row and column, as it is possible to
reach only a finite number of other configurations from a given
configuration within one step, all the same, within one step the given
configuration is reachable only from a finite number of different
configurations.
\end{remark}

\begin{lemma}
\label{l1}
The columns system of a QPA evolution matrix is normalized iff
the condition (\ref{lpc1}), i.e., local probability condition, is satisfied.
\end{lemma}

\begin{lemma}
\label{l2}
The columns system of a QPA evolution matrix is orthogonal iff
the conditions
(\ref{ocv1},\ref{s1a1},\ref{s1b1},\ref{s21},\ref{s3a1},\ref{s3b1}),
i.e., orthogonality of column vectors and separability conditions,
are satisfied.
\end{lemma}

\begin{lemma}
\label{l3}
The rows system of a QPA evolution matrix is normalized iff the condition
(\ref{rvn1}), i.e., row vectors norm condition, is satisfied.
\end{lemma}

\begin{theorem}
\label{th4}
Well-formedness conditions \ref{wfc1} are satisfied iff the evolution 
operator $U_A$ is unitary.
\end{theorem}
\begin{proof}
Lemmas \ref{l1}, \ref{l2}, \ref{l3} imply that Well-formedness conditions
\ref{wfc1} are satisfied iff the columns of the evolution matrix are
orthonormal and rows are normalized. In compliance with Lemma \ref{th3},
columns are orthonormal and rows are normalized iff the matrix is unitary.
\qed
\end{proof}

\begin{remark}
Well-formedness conditions \ref{wfc1} contain the requirement that rows
system has to be normalized, which is not necessary in the case of quantum
Turing machine \cite{BV 97}. Here is taken into account the fact that the
evolution of QPA can violate the unitarity requirement if the row vectors
norm condition is omitted.
\end{remark}

\begin{example}
A QPA, whose evolution matrix columns are orthonormal, however the
evolution is not unitary.
$$
\begin{array}{ll}
Q=\{q\},\ \Sigma=\{1\},\ T=\{1\}.&\\
\delta(q,\#,Z_0,q,\to,Z_01)=1, & \delta(q,\#,1,q,\to,11)=1,\\
\delta(q,1,Z_0,q,\to,Z_01)=1, & \delta(q,1,1,q,\to,11)=1,\\
\delta(q,\$,Z_0,q,\to,Z_01)=1, & \delta(q,\$,1,q,\to,11)=1,
\end{array}
$$
other values of arguments yield $\delta=0$.

By Well-formedness conditions \ref{wfc1}, the columns of the evolution
matrix are orthonormal, but the matrix is not unitary, because the norm
of the rows specified by the configurations $|\omega,Z_0\rangle$ is $0$.
\end{example}

Even in a case of trivial QPA, it is a cumbersome task to check all
the conditions of well-formedness \ref{wfc1}. It is possible to relax the
conditions slightly by introducing a notion of {\em simplified} QPA.

\begin{definition}
\label{def5}
We shall say that a QPA is {\em simplified}, if there exists a function
$D:Q\longrightarrow\{\downarrow,\to\}$, and
$\delta(q_1,\sigma,\tau,q,d,\omega)=0$, if $D(q)\neq d$. Therefore the 
transition function of a simplified QPA is
$$
\varphi(q_1,\sigma,\tau,q,\omega)=\delta(q_1,\sigma,\tau,q,D(q),\omega).
$$
\end{definition}

Taking into account Definition \ref{def5}, following well-formedness 
conditions correspond to simplified QPA:

\begin{wfcond}
\label{wfc2}
\begin{enumerate}
\item Local probability condition.
\begin{eqnarray}
&&\forall(q_1,\sigma_1,\tau_1)\in Q\times\Gamma\times\Delta\nonumber\\
\label{lpc2}
&&\sum\limits_{(q,\omega)\in Q\times\Delta^*} 
|\varphi(q_1,\sigma_1,\tau_1,q,\omega)|^2=1.
\end{eqnarray}
\item Orthogonality of column vectors condition.
\begin{eqnarray}
&&\mbox{For all triples }(q_1,\sigma_1,\tau_1)\neq(q_2,\sigma_1,\tau_2)
\mbox{ in }Q\times\Gamma\times\Delta\nonumber\\
\label{ocv2}
&&\sum\limits_{(q,\omega)\in Q\times\Delta^*}
\varphi^*(q_1,\sigma_1,\tau_1,q,\omega)\varphi(q_2,\sigma_1,\tau_2,q,\omega)=0.
\end{eqnarray}
\item Row vectors norm condition.
\begin{eqnarray}
&&\forall(q_1,\sigma_1,\tau_1,\tau_2)\in Q\times\Gamma\times\Delta^2\nonumber\\
\label{rvn2}
&&\sum\limits_{(q,\tau,\omega)\in Q\times\Delta\times\{\varepsilon,\tau_2,\tau_1\tau_2\}}
|\varphi(q,\sigma_1,\tau,q_1,\omega)|^2=1.
\end{eqnarray}
\item Separability condition.
\begin{eqnarray}
&&\forall(q_1,\sigma_1,\tau_1),(q_2,\sigma_1,\tau_2)\in Q\times\Gamma\times\Delta,
\ \forall\tau_3\in\Delta\nonumber\\
\label{s1a2}
&&a)\ \sum\limits_{(q,\tau)\in Q\times\Delta}
\varphi^*(q_1,\sigma_1,\tau_1,q,\tau)\varphi(q_2,\sigma_1,\tau_2,q,\tau_3\tau)+{}\nonumber\\
&&{}+\sum\limits_{q\in Q}
\varphi^*(q_1,\sigma_1,\tau_1,q,\varepsilon)\varphi(q_2,\sigma_1,\tau_2,q,\tau_3)=0;\\
\label{s1b2}
&&b)\ \sum\limits_{q\in Q}
\varphi^*(q_1,\sigma_1,\tau_1,q,\varepsilon)\varphi(q_2,\sigma_1,\tau_2,q,\tau_2\tau_3)=0.
\end{eqnarray}
\end{enumerate}
\end{wfcond}

\begin{theorem}
The evolution of a simplified QPA is unitary iff Well-formedness conditions
\ref{wfc2} are satisfied.
\end{theorem}
\begin{proof}
By Theorem \ref{th4} and Definition \ref{def5}.\qed
\end{proof}

\section{Language recognition}

Language recognition for QPA is defined as follows. For a QPA\\
$A=(Q,\Sigma,T,q_0,Q_a,Q_r,\delta)$ we define
$C_a=\{|\nu_i q\nu_k,\omega_l\rangle\in C\ |\ q\in Q_a\}$,
$C_r=\{|\nu_i q\nu_k,\omega_l\rangle\in C\ |\ q\in Q_r\}$,
$C_n=C\setminus(C_a\cup C_r)$. $E_a,E_r,E_n$ are subspaces of $H_A$
spanned by $C_a,C_r,C_n$ respectively. We use the observable
$\mathcal{O}$ that corresponds to the orthogonal decomposition
$H_A=E_a\oplus E_r\oplus E_n$. The outcome of each observation is
either ``accept" or ``reject" or ``non-halting".

The language recognition is now defined as follows: For an 
$x\in\Sigma^*$ we consider as an input $\#x\$$, and assume that
the computation starts with $A$ being in the configuration
$|q_0\#x\$,Z_0\rangle$. Each computation step consists of two parts. At
first the linear operator $U_A$ is applied to the current global state
and then the resulting superposition is observed using the observable
$\mathcal{O}$ as defined above. If the global state before the observation
is $\sum\limits_{c\in C}\alpha_c|c\rangle$, then the probability that the
resulting superposition is projected into the subspace $E_i$,
$i\in\{a,r,n\}$, is $\sum\limits_{c\in C_i}|\alpha_c|^2$. The
computation continues until the result of an observation is ``accept" or
``reject".

\begin{definition}
We shall say that an automaton is a deterministic reversible pushdown
automaton (RPA), if it is a simplified QPA with
$\varphi(q_1,\sigma,\tau,q,\omega)\in\{0,1\}$ and there exists a function 
$f:Q\times\Gamma\times\Delta\longrightarrow Q\times\Delta^*$, such that
$f(q_1,\sigma,\tau)=(q,\omega)$ if and only if $\varphi(q_1,\sigma,\tau,q,\omega)=1$.
\end{definition}

We can regard $f$ as a transition function of a RPA.
Needless to say, if any language is recognized by a RPA, it is recognized
with probability equal to $1$. Note that the local probability condition (\ref{lpc2}) is satisfied
automatically for RPA.

\begin{theorem}
\label{RegularLanguages}
Every regular language is recognizable by some QPA.
\end{theorem}
\begin{proof}
It is sufficient to prove that any deterministic finite automaton (DFA) can be simulated by RPA.
Let us consider a DFA with $n$ states $A_{DFA}=(Q_{DFA}, \Sigma, q_0, Q_F, \delta)$, where 
$\delta:Q_{DFA}\times\Sigma\longrightarrow Q_{DFA}$.

To simulate $A_{DFA}$ we shall construct a RPA $A_{RPA}=(Q,\Sigma,T,q_0,Q_a,Q_r,\varphi)$ with the number of states $2n$.

The set of states is $Q=Q_{DFA}\cup Q'_{DFA}$, where $Q'_{DFA}$ are the newly introduced states, which are linked to
$Q_{DFA}$ by a one-to-one relation $\{(q_i,q'_i)\in Q_{DFA}\times Q'_{DFA}\}$. Thus $Q_F$ has one-to-one relation to
$Q'_F\subset Q'_{DFA}$.

The stack alphabet is $T=Ind(Q_{DFA})$, where $\forall i\ Ind(q_i)=i$;
the set of accepting states is $Q_a=Q'_F$ and the set of rejecting states is
$Q_r=Q'_{DFA}\setminus Q'_F$. As for the function $D$, $D(Q_{DFA})=\{\to\}$ and $D(Q'_{DFA})=\{\downarrow\}$.

We shall define sets $R$ and $\overline{R}$ as follows:
$$
R=\{(q'_j,\sigma,i)\in Q'_{DFA}\times\Sigma\times T\ |\ \delta(q_i,\sigma)=q_j\};
$$
$$
\overline{R}=\{(q'_j,\sigma,i)\in Q'_{DFA}\times\Sigma\times T\ |\ \delta(q_i,\sigma)\neq q_j\}.
$$
The construction of the transition function $f$ is performed by the following rules:
\begin {enumerate}
\item $\forall (q_i,\sigma,\tau)\in Q_{DFA}\times\Sigma\times\Delta\ \ f(q_i,\sigma,\tau)=(\delta(q_i,\sigma),\tau i)$;
\label{rule1}
\item $\forall (q'_j,\sigma,i)\in R\ \ f(q'_j,\sigma,i)=(q'_i,\varepsilon)$;
\label{rule2}
\item $\forall (q'_j,\sigma,i)\in \overline{R}\ \ f(q'_j,\sigma,i)=(q_j,i)$;
\label{rule3}
\item $\forall (q'_j,\sigma)\in Q'_{DFA}\times\Sigma\ \ f(q'_j,\sigma,Z)=(q_j,Z)$;
\label{rule4}
\item $\forall (q,\tau)\in Q\times\Delta\ \ f(q,\#,\tau)=(q,\tau)$;
\label{rule5}
\item $\forall (q_i,\tau)\in Q_{DFA}\times\Delta\ \ f(q_i,\$,\tau)=(q'_i,\tau)$;
\label{rule6}
\item $\forall (q'_i,\tau)\in Q'_{DFA}\times\Delta\ \ f(q'_i,\$,\tau)=(q_i,\tau)$.
\label{rule7}
\end{enumerate}

Thus we have defined $f$ for all the possible arguments. Our automaton simulates the DFA. Note that the automaton may
reach a state in $Q'_{DFA}$ only by reading the end-marking symbol $\$$ on the input tape. As soon as $A_{RPA}$ reaches the
end-marking symbol $\$$, it goes to an accepting state, if its current state is in $Q_F$, and goes to a rejecting state
otherwise.

The construction is performed in a way so that $A_{RPA}$ satisfies Well-formedness conditions \ref{wfc2}.

As we know, RPA automatically satisfies the local probability condition (\ref{lpc2}).

Let us prove, that the automaton satisfies the orthogonality condition (\ref{ocv2}).

For RPA, the condition (\ref{ocv2}) is equivalent to the requirement that for all triples
$(q_1,\sigma_1,\tau_1)\neq(q_2,\sigma_1,\tau_2)\ \ f(q_1,\sigma_1,\tau_1)\neq f(q_2,\sigma_1,\tau_2)$.

If $q_1,q_2\in Q_{DFA}$, $f(q_1,\sigma_1,\tau_1)\neq f(q_2,\sigma_1,\tau_2)$ by rule \ref{rule1}.

Let us consider the case when $(q_1,\sigma_1,\tau_1),(q_2,\sigma_1,\tau_2)\in R$. We shall denote $q_1,q_2$ as
$q'_i,q'_j$ respectively. Let us assume from the contrary that $f(q'_i,\sigma_1,\tau_1)=f(q'_j,\sigma_1,\tau_2)$. 
By rule \ref{rule2}, $(q'_{\tau_1},\varepsilon)=(q'_{\tau_2},\varepsilon)$. Hence $\tau_1=\tau_2$. By the definition of $R$,
$\delta(q_{\tau_1},\sigma_1)=q_i$ and $\delta(q_{\tau_2},\sigma_1)=q_j$. Since $\tau_1=\tau_2$, $q_i=q_j$. Therefore 
$q'_i=q'_j$, i.e., $q_1=q_2$. We have come to a contradiction with the fact that 
$(q_1,\sigma_1,\tau_1)\neq(q_2,\sigma_1,\tau_2)$.

If $(q_1,\sigma_1,\tau_1),(q_2,\sigma_1,\tau_2)\in \overline{R}$, $f(q_1,\sigma_1,\tau_1)\neq f(q_2,\sigma_1,\tau_2)$
by rule \ref{rule3}.

If $q_1\in Q_{DFA},q_2\in Q'_{DFA}$ then $f(q_1,\sigma_1,\tau1)\neq f(q_2,\sigma_1,\tau_2)$ by rules \ref{rule1}, 
\ref{rule2}, \ref{rule3}.

In case $\tau_1$ or $\tau_2$ is $Z$, or $\sigma_1\in\{\#,\$\}$, proof is straightforward.

The compliance with row vectors norm condition (\ref{rvn2}) and separability conditions (\ref{s1a2}) and (\ref{s1b2}) is
proved in the same way. \qed
\end{proof}

\begin{example}
First, let us consider a language $L_1=(0,1)^*1$, for which we know that it
is not recognizable by QFA \cite{KW 97}.

Language $L_1$ is recognizable by a RPA.
Let us consider a deterministic finite automaton with two states $q_0,q_1$ and the 
following transitions: $\delta (q_0,0)=q_0$, $\delta (q_0,1)=q_1$,
$\delta (q_1,0)=q_0$, $\delta(q_1,1)=q_1$.\\
It is possible to transform this automaton to a RPA, which
satisfies the corresponding well-formedness conditions:\\
$Q=\{q_0,q_1,q_2,q_3,q_4,q_5\}$,
$Q_a=\{q_5\}$, $Q_r=\{q_4\}$,
$\Sigma=\{0,1\}$,
$T=\{0,1\}$.\\
The states $q_0,q_1$ have the same semantics as in the deterministic
prototype, the only difference is in case input tape symbols $0$
or $1$ is read, when each transition starting in the state $q_0$,
automaton pushes $0$ into stack, whereas in the state $q_1$ pushes $1$.
After reaching the endmarking symbol $\$$, depending on its current state, 
the automaton goes to the state $q_5$ or $q_6$.\\
Finally, we have to add two more states $q_2,q_3$ to our RPA, to ensure 
its unitarity. Values of the transition function follow:
$$
\begin{array}{lll}
\forall\tau\in\Delta\ \forall q\in Q\ \forall\sigma\in\Sigma;&\\
\varphi(q,\#,\tau,q,\tau)=1, & &\\
\varphi(q_0,0,\tau,q_0,\tau 0)=1, & \varphi(q_1,0,\tau,q_0,\tau 1)=1, & D(q_0)=\to,\\
\varphi(q_0,1,\tau,q_1,\tau 0)=1, & \varphi(q_1,1,\tau,q_1,\tau 1)=1, & D(q_1)=\to,\\
\varphi(q_0,\$,\tau,q_4,\tau)=1, & \varphi(q_1,\$,\tau,q_5,\tau)=1, & D(q_2)=\downarrow,\\
\varphi(q_2,1,\tau,q_0,\tau)=1, & \varphi(q_3,0,\tau,q_1,\tau)=1, & D(q_3)=\downarrow,\\
\varphi(q_2,\$,\tau,q_2,\tau)=1, & \varphi(q_3,\$,\tau,q_3,\tau)=1, & D(q_4)=\downarrow,\\
\varphi(q_4,\sigma,\tau,q_4,\tau)=1, & \varphi(q_5,\sigma,\tau,q_5,\tau)=1, & D(q_5)=\downarrow,\\
\varphi(q_2,0,Z,q_0,Z)=1, & \varphi(q_3,1,Z,q_1,Z)=1, & D(q_6)=\downarrow,\\
\varphi(q_2,0,0,q_2,\varepsilon)=1, & \varphi(q_2,0,1,q_3,\varepsilon)=1, &\\
\varphi(q_3,1,0,q_2,\varepsilon)=1, & \varphi(q_3,1,1,q_3,\varepsilon)=1, &\\
\varphi(q_4,\$,\tau,q_0,\tau)=1, & \varphi(q_5,\$,\tau,q_1,\tau)=1, &
\end{array}
$$
other values of arguments yield $\delta=0$.\\
Let us note that states $q_2,q_3$ are not reachable from the initial state
$q_0$, however they are necessary to make the automaton unitary.\qed
\end{example}

Let us consider a language which is not regular, namely,
$$L_2=\{\omega\in(a,b)^*|\ |\omega|_a=|\omega|_b\},$$ where $|\omega|_i$
denotes the number of occurrences of the symbol $i$ in the word $\omega$.

\begin{lemma}
\label{LemmaL3}
Language $L_2$ is recognizable by a RPA.
\end{lemma}
\begin{proof}
Our RPA has four states $q_0,q_1,q_2,q_3$, where $q_2$ is an accepting
state, whereas $q_3$ - rejecting one. Stack alphabet $T$ consists of
two symbols $1,2$. Stack filled with 1's means that the processed part
of the word $\omega$ has more occurrences of a's than b's, whereas 2's
means that there are more b's than a's. Furthermore, length of the
stack word is equal to the difference of number of a's and b's. Empty stack
denotes that the number of a's and b's is equal.

Values of the transition function follow:
$$
\begin{array}{lll}
\forall q\in Q\ \forall\tau\in\Delta; & &\\
\varphi(q,\#,\tau,q,\tau)=1, & \varphi(q_0,a,Z,q_0,Z1)=1, & D(q_0)=\to,\\
\varphi(q_0,b,Z,q_0,Z2)=1, & \varphi(q_0,\$,Z,q_2,Z1)=1, & D(q_1)=\downarrow,\\
\varphi(q_0,a,1,q_0,11)=1, & \varphi(q_0,b,1,q_1,\varepsilon)=1, & D(q_2)=\downarrow,\\
\varphi(q_0,\$,1,q_3,1)=1, & \varphi(q_0,a,2,q_1,\varepsilon)=1, & D(q_3)=\downarrow,\\
\varphi(q_0,b,2,q_0,22)=1, & \varphi(q_0,\$,2,q_3,2)=1, & \varphi(q_1,a,Z,q_0,Z)=1,\\
\varphi(q_1,b,Z,q_0,Z)=1, & \varphi(q_1,\$,\tau,q_1,\tau)=1, & \varphi(q_1,a,1,q_3,12)=1,\\
\varphi(q_1,b,1,q_0,1)=1, & \varphi(q_1,a,2,q_0,2)=1, & \varphi(q_1,b,2,q_3,21)=1,\\
\varphi(q_2,a,Z,q_3,Z2)=1, & \varphi(q_2,b,Z,q_3,Z1)=1, & \varphi(q_2,\$,Z,q_0,Z)=1,\\
\varphi(q_2,a,1,q_2,\varepsilon)=1, & \varphi(q_2,b,1,q_0,12)=1, & \varphi(q_2,\$,1,q_0,1)=1,\\
\varphi(q_2,a,2,q_0,21)=1, & \varphi(q_2,b,2,q_2,\varepsilon)=1, & \varphi(q_2,\$,2,q_0,2)=1,\\
\forall\sigma\in\{a,b,\$\} & \varphi(q_3,\sigma,Z,q_3,Z)=1, &\\
\varphi(q_3,a,1,q_3,1)=1, & \varphi(q_3,b,1,q_3,11)=1, & \varphi(q_3,\$,1,q_2,1)=1,\\
\varphi(q_3,a,2,q_3,22)=1, & \varphi(q_3,b,2,q_3,2)=1, & \varphi(q_3,\$,2,q_2,2)=1,
\end{array}
$$
other values of arguments yield $0$.
\end{proof}\qed

Let us consider language which is not recognizable by any deterministic
pushdown automaton:

\begin{theorem}
Language $L_3=\{\omega\in(a,b,c)^*|\ |\omega|_a=|\omega|_b=|\omega|_c\}$ is
recognizable by a QPA with probability $\frac{2}{3}$.
\end{theorem}
\begin{proof}
Sketch of proof. The automaton takes three equiprobable actions, during the
first action it compares $|\omega|_a$ to $|\omega|_b$, whereas during the
second action $|\omega|_b$ to $|\omega|_c$ is compared. Input word is
rejected if the third action is chosen. Acceptance probability totals
$\frac{2}{3}$.\end{proof}\qed

\begin{theorem}
Language
$L_5=\{\omega\in(a,b,c)^*|\ |\omega|_a=|\omega|_b\mbox{ xor }|\omega|_a=|\omega|_c\}$
is recognizable by a QPA with probability $\frac{4}{7}$.
\end{theorem}
\begin{proof}
Sketch of proof. The automaton starts the following actions with the
following amplitudes:\\
a) with an amplitude $\sqrt{\frac{2}{7}}$ compares $|\omega|_a$ to 
$|\omega|_b$.\\
b) with an amplitude $-\sqrt{\frac{2}{7}}$ compares $|\omega|_a$ to
$|\omega|_c$.\\
c) with an amplitude $\sqrt{\frac{3}{7}}$ accepts the input.
If exactly one comparison gives positive answer, input is accepted with
probability $\frac{4}{7}$. If both comparisons gives positive answer,
amplitudes, which are chosen to be opposite, annihilate and the input is
accepted with probability $\frac{3}{7}$.\end{proof}\qed

Language $L_5$ cannot be recognized by deterministic pushdown automata.

An open problem is to find a language, not recognizable by probabilistic pushdown automata as well.

\end{document}